\documentclass{radioeng_modified}
\usepackage[czech,english]{babel}

\usepackage[cp1250]{inputenc}
\usepackage{graphicx} % To include and scale graphics
\usepackage{color,soul}
\usepackage{cite}
\usepackage{amsmath,bm}

\usepackage{amssymb}
\usepackage{tabularx}
\usepackage{textcomp}

\usepackage{listings}  %To include listings

\begin{document}

\setcounter{firstpage}{1}

\rengHeader{}{}{
... %1
%JUNE %2
%SEPTEMBER %3
%DECEMBER %4
}{Y. V. PERSHIN, SPICE MODELING OF MEMCOMPUTING LOGIC GATES}

\rengTitle{SPICE Modeling of Memcomputing Logic Gates}

\rengNames{Yuriy~V.~PERSHIN $^\mathit{1}$}

\rengAffil{$^1$ Department of Physics and Astronomy, University of South Carolina, Columbia, SC 29208 USA}

\rengMail{pershin@physics.sc.edu}

\rengReceived{...}{...} % submitted accepted (do not fill out)

%%%%%%%%%%%%%%%%%%%%%%%%%%%%%%%%%%%%%%%%%%%%%%%%%%%%%%%%%%%%%%
% Abstract, keywords
%%%%%%%%%%%%%%%%%%%%%%%%%%%%%%%%%%%%%%%%%%%%%%%%%%%%%%%%%%%%%%%
\begin{multicols}{2}

\begin{rengAbstract}
Memcomputing logic gates generalize the traditional Boolean logic gates for operation in the reverse direction. According to the literature, this functionality enables the efficient solution of computationally-intensive problems including factorization and NP-complete problems. To approach the deployment of memcomputing gates in hardware, this paper introduces SPICE models of memcomputing logic gates following their original definition. Using these models, we demonstrate the behavior of single gates as well as small self-organizing circuits. We also correct some inconsistencies in the prior literature. Importantly, the correct schematics of dynamic correction module is reported here for the first time. Our work makes memcomputing more accessible to those who are interested in this emerging computing technology.
\end{rengAbstract}

\rengKeywords{Memristors, SPICE, nonlinear dynamical systems, computing technology}

\rengSection{Introduction}

Digital memcomputing machines are an emerging class of unconventional computing systems developed to efficiently solve factorization and combinatorial optimization problems~\cite{Traversa17a,MemComputingbook}. Fundamentally, these are complex dynamical systems with deterministic continuous dynamics whose phase space contains an attractor corresponding to the problem solution (or multiple attractors if several solutions are possible). According to Traversa and Di Ventra, when realized in hardware, 
digital memcomputing machines offer a polynomial-time solution to the factorization and NP-complete problems~\cite{Traversa17a}. It has been argued that the dynamics of digital memcomputing machines is deterministic, non-chaotic, and without periodic orbits~\cite{no-chaosa,no-chaosb}. Moreover, the operation of these machines is topologically robust against perturbations and noise~\cite{topo,DMtopo}.  So far the research on digital memcomputing has been substantially focused on software simulations of ordinary differential equations representing the circuit dynamics.

\begin{figure}
 \begin{center}
  \includegraphics[width=0.8\columnwidth,keepaspectratio]{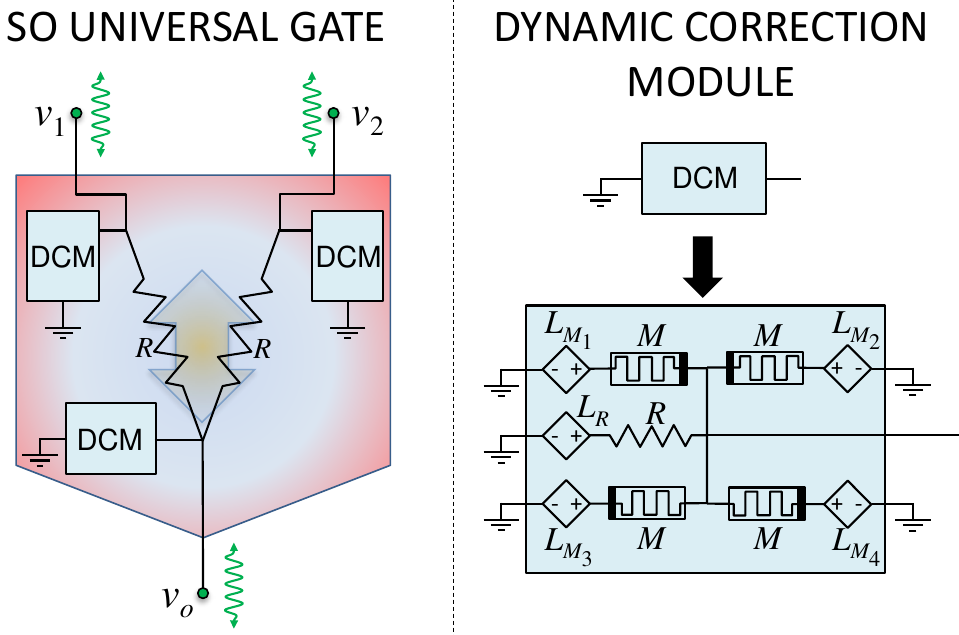}
   \fcaption{Left panel: the universal self-organizing logic gate is composed of three dynamic correction modules (DCMs) and two resistors. Right panel: internal structure of the dynamic correction module (incorrect). Here, the resistive memories M~\cite{chua76a,09_memelements,11_memory_materials} have minimum and maximum resistances $R_{on}$ and $R_{off}$, respectively, and the resistor's resistance $R=R_{off}$.  $L_{M_j}$-s and $L_R$ are voltage-controlled voltage generators.  As we explain in the text, in the right panel, the polarity of all memristive elements must be reversed. Reprinted from Ref.~\cite{Traversa17a}.}
   \label{fig:1}
  \end{center}
\end{figure}

Three designs of memcomputing logic gates are available in the literature. The most intricate is the original design~\cite{Traversa17a} (Design I) that is presented in Fig.~\ref{fig:1}. According to Fig.~\ref{fig:1}, self-organizing AND, OR or XOR can be built using twelve memristive elements, fifteen voltage-controlled voltage generators, and several resistors. Moreover, some auxiliary circuitry is required to ensure that the final states of these gates (operating in the continuous or analog domain) are binary. A simplified design of self-organizing AND and OR (Design II) was introduced by Bearden et al.~\cite{Bearden} wherein each gate requires five memristive elements, six voltage generators, and few resistors. %In this approach, the auxiliary circuitry is also required. 
In principle, the simplified gates can perform the same tasks as the original ones. A study shows that Design II gates can be built using physical memristive devices~\cite{Ochs20a}. In the third design (Design III), the gates  are defined by a set of differential equations~\cite{Sean3SAT}. In fact, Design III can be considered as a variation of analogSAT~\cite{zoltan,molnar2020accelerating} that does not suffer from the exponential fluctuations in the energy function~\cite{Sean3SAT}.
 
The purpose of this work is to develop SPICE models of Design I self-organizing logic gates~\cite{Traversa17a}. During the last decade or so, the SPICE modeling of adaptive circuit elements (known as memrisitive, memcapacitive and meminductive systems~\cite{chua76a, 09_memelements,Membook})  has become increasingly important and resulted in various SPICE models of deterministic memelements (see, for instance, Refs.~\cite{Biolek2009-2,pershin13c,Biolek13a,da2014two,vourkas2015spice,li2015memristor,Biolek16a,garcia2016spice,schroedter2022spice}). A notable recent development is the simulation of probabilistic memristive devices in SPICE~\cite{dowling2022analytic}. SPICE is a general-purpose circuit simulation program~\cite{vladimirescu1994spice,kundert2006designer}. In SPICE, the circuit can be first built graphically and then simulated numerically using pre-defined or user-defined models of individual circuit components. Such user-defined models of Design I self-organizing logic gates~\cite{Traversa17a} are formulated in this paper.

In this work, we used LTSpice~XVII (Analog Devices) as the simulation tool. Our SPICE models may need minor adjustments to be used on other SPICE simulators (e.g., PSPICE, Ngspice). 

The paper is organized as follows. We start with the
 introduction of Design I self-organizing logic gates (Sec.~\ref{sec:2}) that is followed by a presentation of their SPICE models (Sec.~\ref{sec:3}). In Sec.~\ref{sec:4}, we give examples of SPICE simulations of individual gates and  circuits thereof. Sec.~\ref{sec:5} concludes the paper. Complete listings of LTspice codes are given  in the Appendix B. 
 %We expect that the SPICE models presented in this manuscript will be helpful to students and researchers interested in emerging computing approaches.

\rengSection{Self-organizing gates and circuits} \label{sec:2}

Self-organizing logic gates generalize the traditional logic gates for operation in the reverse direction~\cite{Traversa17a,MemComputingbook}. In these gates, each terminal serves the double function of input and output. Although self-organizing gates operate in the analog mode, the auxiliary circuitry (voltage-controlled differential current generators presented below) as well as the gate design ensure that the final states are binary. In what follows, Boolean 1 and 0 are represented by $v_c=1$~V and $-v_c=-1$~V voltage levels, respectively.

The logic behind the construction of self-organizing gates can be partially captured from the following excerpt from~\cite{Traversa17a}: ``if the gate is connected to a network and the gate configuration
is correct, no current flows from any terminal (the gate is in stable equilibrium). Otherwise, a current of the order of $v_c/R_{on}$ flows with sign
opposite to the sign of the voltage at the terminal.'' Below, this property of self-organizing gates is used to demonstrate that the polarity of the memristive elements in Fig.~\ref{fig:1} circuit must be reversed for the correct gate operation.

The transformation of traditional Boolean logic circuits into Design I self-organizing logic circuits~\cite{Traversa17a}
involves the following steps:
\begin{itemize}
    \item Replacing the traditional logic gates with self-organizing gates of the same type.
    \item Representing  the external input signals by constant-value voltage sources.
    \item Adding auxiliary circuitry: A voltage-controlled differential current generator (VCDCG) is added to each node but the nodes that are used for input signals. By node, we mean either the point of connection of two or more gate terminals or an unconnected gate terminal. 
    \item The external input signals are applied at the initial moment of time and stay constant with time. The end of dynamics indicates that a solution is found and can be read. The infinite dynamics implies the absence of a solution.
\end{itemize}
For the sake of completeness, next, we provide the minimal description of the circuit components in the Design I self-organizing logic circuits that is required for their implementation in SPICE. These definitions were extracted from Refs.~\cite{Traversa17a,di2018self,MemComputingbook}. 

\rengSubsection{Voltage-controlled voltage generators} \label{sec2_1}

Let us consider the structure of the  universal self-organizing gate shown in Fig.~\ref{fig:1}. 
Its ultimate functionality (e.g., self-organizing AND, OR, or XOR) is defined by the equations governing the  voltage-controlled voltage generators (VCVGs)
$L_{M_1}-L_{M_4}$ and $L_R$. According to Ref.~\cite{Traversa17a}, the voltage across VCVG is a linear function of the gate voltages
\begin{equation}
v_{VCVG}=a_1v_1+a_2v_2+a_ov_o+dc,
\end{equation}

%\vspace{-0.3cm}
\noindent where $v_1$, $v_2$, and $v_0$ are the gate voltages, and  $a_1$, $a_2$, $a_0$, and $dc$ are the constants. For the sake of convenience, these constants are specified in Table~\ref{tbl:1} (Appendix A). 

\rengSubsection{Memristive elements}

Having defined the voltage-controlled voltage generators, next, we consider the memristive elements M in Fig.~\ref{fig:1}. The  response of the memristive system $j$  is described by Ohm's law
\begin{eqnarray} \label{eq:2}
v_{M_j}(t)= M(x_j)i_{M_j}(t),
\end{eqnarray}
where $v_{M_j}$ is the voltage (defined with respect to the thick-bar terminal of the circuit symbol of M), $i_{M_j}$ is the current, 
\begin{equation}
     M(x_j)=\left( R_{off}-R_{on} \right)x_j +R_{on}
\end{equation}

\vspace{-0.3cm}
\noindent is the state-dependent resistance (memristance),
$x_j\in [0,1]$ is the internal state variable~\cite{chua76a}, 
$R_{on}$ and $R_{off}$ are the on- and off-state resistances. 

The dynamics of $x_J$ follows the ordinary differential equation
\begin{equation} \label{eq:x}
    \frac{\textnormal{d}x_j}{\textnormal{d}t}=-\alpha h(x_j,v_{M_j})\left[
\left( R_{off}-R_{on} \right)x_j +R_{on} \right]^{-1} v_{M_j},
\end{equation}

\vspace{-0.1cm}
\noindent where, $\alpha$ is the constant and the function $h(x,v_M)$
influences the dynamics of the internal state. While many choices for 
$h(x,v_M)$ are available, following~\cite{Traversa17a}, we use 
\begin{equation}
h(x,v_M)=\theta\left(x\right)\theta\left(v_M\right)+
\theta\left(1-x\right)\theta\left(-v_M\right), \label{eq:h}
\end{equation} \vspace{0.4cm}
\noindent where $\theta (...)$ is the  unit step function. According to 
Eqs.~(\ref{eq:x}) and (\ref{eq:h}),  $x_j$ decreases down to 
 $x_j=0$ at positive voltages. At negative voltages, $x_j$ increases up to $x_j=1$.
It is not difficult to recognize that
Eqs.~(\ref{eq:2}), (\ref{eq:x}), and (\ref{eq:h}) correspond to the
ideal memristor model~\cite{08_strukov}. The limitations of this model are well known~\cite{Membook}. Note that Table~\ref{tbl:params} (Appendix A) lists the values of parameters used in the prior simulations~\cite{Traversa17a}. Eq.~(\ref{eq:h}) corresponds to $V_t=0$ and $k=\infty$.  For definition of $V_t$ and $k$, see ~\cite{Traversa17a}.

Moreover, a parasitic device capacitance is taken into account using a constant-value capacitor of capacitance $C$ connected 
in parallel to every memrstive element. These capacitors are not shown
in Fig.~\ref{fig:1} explicitly but taken into account in the model. 

\begin{figure}
 \begin{center}
  (a)\includegraphics[width=0.8\columnwidth,keepaspectratio]{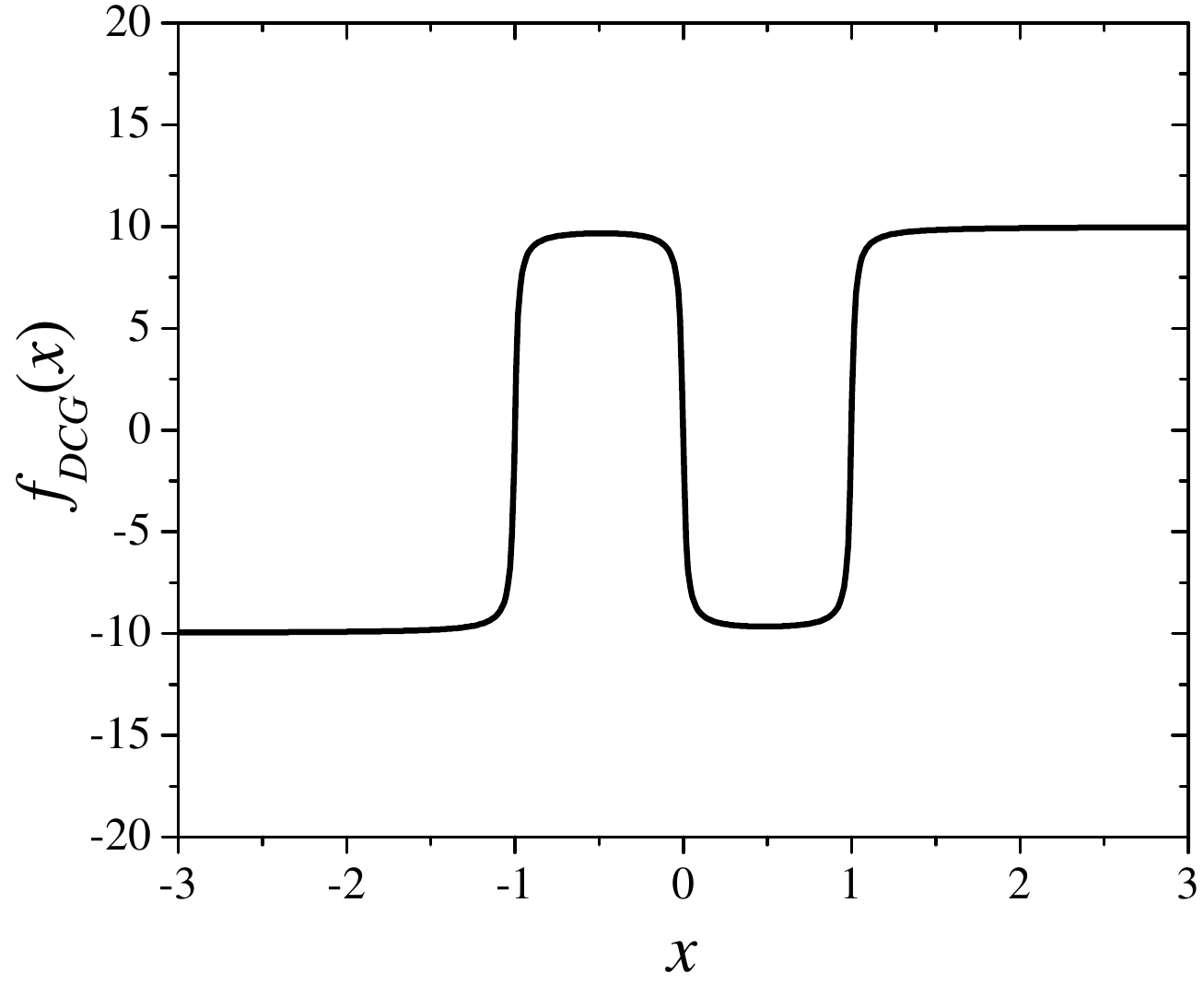} \\
   (b) \includegraphics[width=0.8\columnwidth,keepaspectratio]{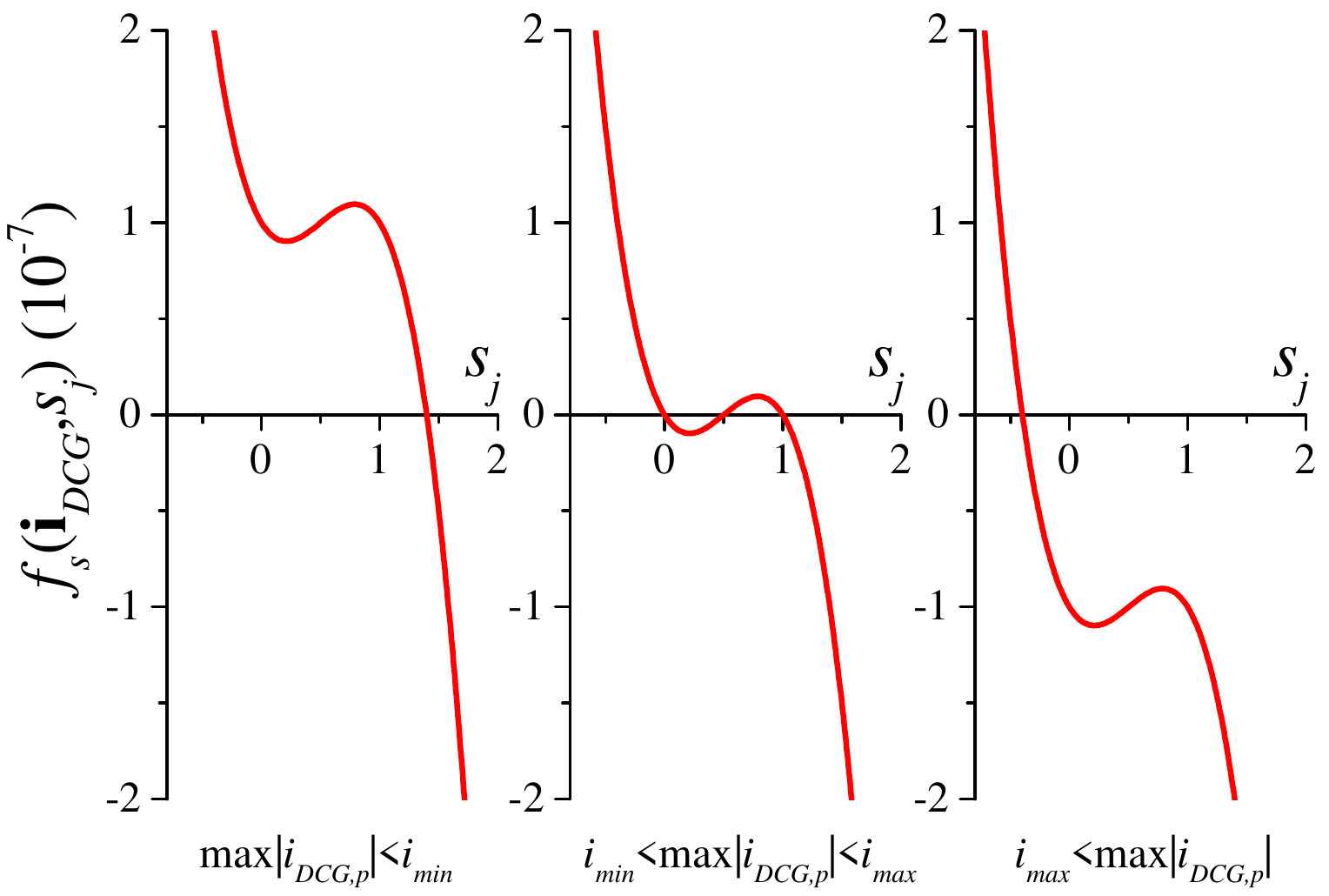}
   \fcaption{Functions (a)  $f_{DCG}(x)$  and (b) 
   $f_s\left( \mathbf{i}_{DCG},s_j\right)$ defined by 
   Eqs.~(\ref{eq:fDCG}) and (\ref{eq:8}), respectively. These graphs were obtained using parameters values from Table~\ref{tbl:params} (Appendix A).}
   \label{fig:functions}
  \end{center}
\end{figure}

\rengSubsection{Voltage-controlled differential current generators} \label{sec:VCDCG}
\vspace{0.3 cm}

Finally, we introduce equations describing the voltage-controlled differential current generators. These generators are second-order dynamical systems whose evolution follows~\cite{Traversa17a,di2018self}:
\begin{eqnarray}
    \nonumber \frac{\text{d}i_{DCG,j}}{\text{d}t}&=&\theta\left(s_j-\frac{1}{2} \right)f_{DCG}(v_{DCG,j})- \\ & &
     \hspace{1.5cm} \gamma\theta\left(\frac{1}{2}-s_j\right)i_{DCG,j}, \label{eq:5} \\
     \frac{\text{d}s_j}{\text{d}t}&=&f_s\left( \mathbf{i}_{DCG},s_j\right). \label{eq:6}
\end{eqnarray}
Here,  $\gamma$ is the constant, $v_{DCG,j}$ is the voltage at the node that VCDCG is connected to, $i_{DCG,j}$ is the current, $\mathbf{i}_{DCG}$ is the vector of the currents of all VCDCGs, and $s_j$ is the second state variable. Note that the above equation corresponds to $\delta s=0$ in Table~\ref{tbl:params} (Appendix A).  For definition of $\delta s$, see ~\cite{Traversa17a}. Our specific realization of $f_{DCG}(x)$ in Eq.~(\ref{eq:5}) is based on the
 arctangent functions (one of the suggested realizations of $ f_{DCG}(x)$~\cite{Traversa17a}). Specifically, we use
\begin{eqnarray}
    f_{DCG}(x)=\frac{2q}{\pi}\left(\text{arctan}\left[\frac{m_1\pi}{2q}\left(x+v_c\right)\right]+\right. \hspace{1.0cm} \label{eq:fDCG} && \\
    \text{arctan}\left[\frac{m_0\pi}{2q}x\right]+ 
    \left. \text{arctan}\left[\frac{m_1\pi}{2q}\left(x-v_c\right)\right]    \right) , && \; \nonumber
\end{eqnarray}
where $q$, $m_0$, $m_1$, and $v_c$ are the constants. Eq.~(\ref{eq:fDCG}) is illustrated in Fig.~\ref{fig:functions}(a).

In Eq.~(\ref{eq:6}), the function $f_s\left( \mathbf{i}_{DCG},s_j\right)$ is defined as
\begin{eqnarray} \label{eq:8}
    f_s\left( \mathbf{i}_{DCG},s_j\right)=-k_ss_j(s_j-1)(2s_j-1)- \hspace{2.0cm} &&  \\
   k_i\left(  1-\prod_p \theta\left( i_{min}^2-i_{DCG,p}^2 \right) 
     -\prod_p \theta\left( i_{max}^2-i_{DCG,p}^2 \right) \right).  && \nonumber
\end{eqnarray}
Here, $k_s$, $k_i$, $i_{min}$, and $i_{max}$ are the positive constants ($i_{min}<i_{max}$). Eq.~(\ref{eq:8}) corresponds to $\delta i=0$ in Table~\ref{tbl:params} (Appendix A).  For definition of $\delta i$, see ~\cite{Traversa17a}.

Importantly, unlike Refs.~\cite{Traversa17a,di2018self} we use the minus sign in front of $k_i$ term in Eq.~(\ref{eq:8}). 
The minus sign is required to implement the following anticipated purpose of 
 $s_j$-s: the reset of all $|i_{DCG,j}|$ to below $i_{min}$ as soon as at least one of $|i_{DCG,j}|$ exceeds $i_{max}$.

Eq.~(\ref{eq:8}) is illustrated in Fig.~\ref{fig:functions}(b). 
Fig.~\ref{fig:functions}(b) (left panel) shows that when the absolute value of all currents is less than $i_{min}$, the function $f_s\left( \mathbf{i}_{DCG},s_j\right)$ has a single zero at $s>1$, which is a stable fixed point of Eq.~(\ref{eq:6}). When the absolute value of at least one $i_{DCG,p}$ is larger than $i_{max}$, the stable fixed point is located at $s<0$, see Fig.~\ref{fig:functions}(b) (right panel). In the intermediate case, there are two stable and one unstable fixed point (Fig.~\ref{fig:functions}(b) (middle panel)). Therefore, when the absolute value of one of the currents exceeds $i_{max}$, all variables $s_j$ -- that are described by the identical equations -- start drifting towards the negative stable fixed point, causing the relaxation of VCDCG currents (through the last term in Eq.~(\ref{eq:5})).  The normal response of VCDCGs (due to the first term in Eq.~(\ref{eq:5})) is restored later, after the condition  $|i_{DCG,j}|<i_{min}$ for all $j$ is satisfied. 

\rengSection{SPICE models} \label{sec:3}

\rengSubsection{Basic details}

In our SPICE models we have attempted to utilize the parameters' values used in Ref.~\cite{Traversa17a} (Table~\ref{tbl:params}, Appendix A). SI units are assumed even though this assumption can not be completely justified. For instance, $R_{on}=0.01$~$\Omega$ and $R_{off}=1$~$\Omega$ are much smaller than the on- and off-state resistances observed in experimental devices~\footnote{Typically, these resistances range from kiloohms to megaohms.}. 

Appendix B contains LTspice models of self-organizing AND, OR, and XOR, memristive elements, and voltage-controlled differential current generators (Tables~\ref{tbl:SAND}-\ref{tbl:VCDCG}). These models were formulated following the common practices in SPICE modeling~\cite{Biolek13a}. For instance, to integrate the differential equations (\ref{eq:x}), (\ref{eq:5}), and (\ref{eq:6}), we use capacitors that are charged or discharged with voltage-controlled current sources representing the right-hand sides of these equations, etc.

\begin{figure}
 \begin{center}
  \includegraphics[width=0.8\columnwidth,keepaspectratio]{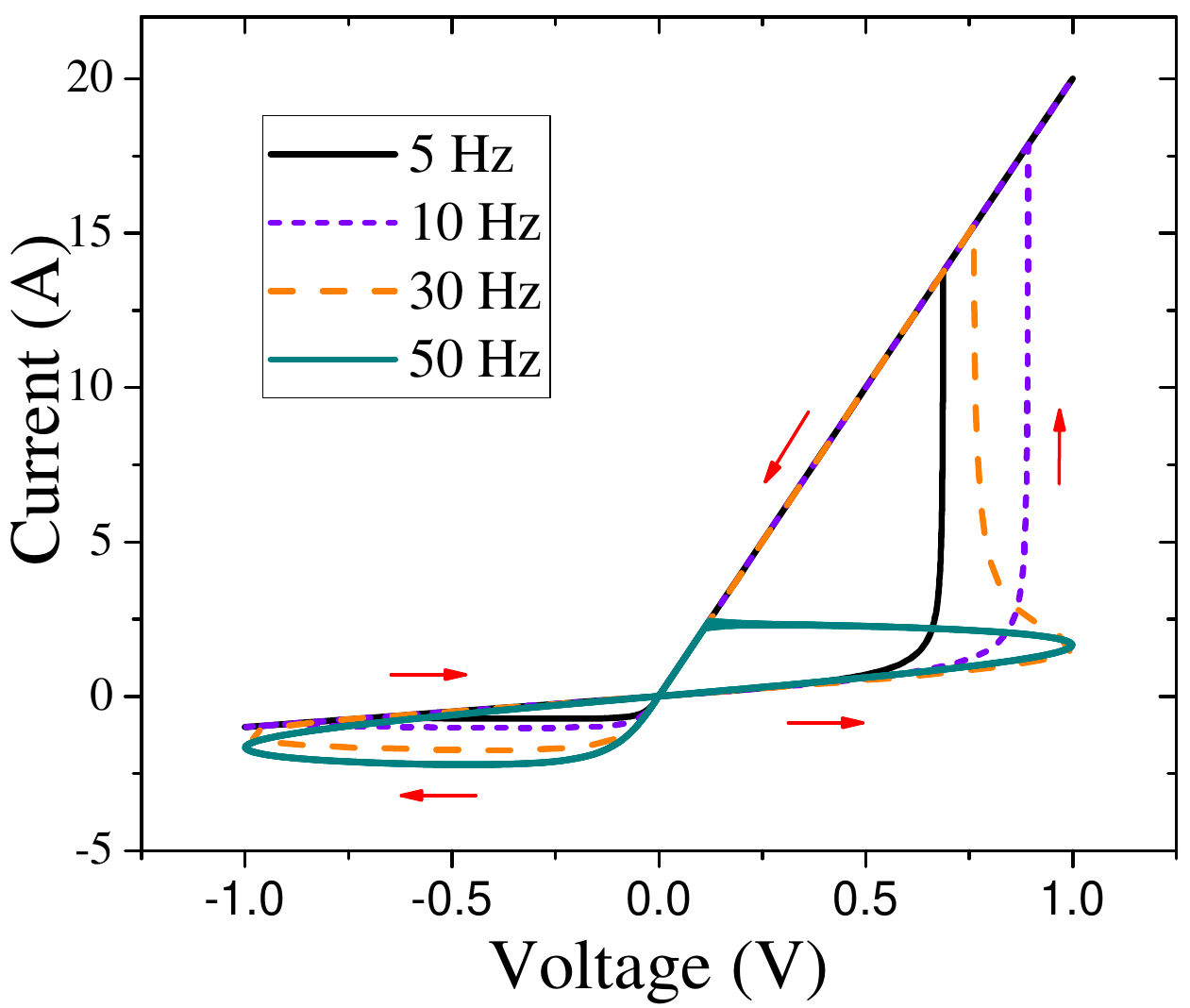} 
   \fcaption{Current-voltage curves of the memristive element subjected to a sinusoidal voltage. These curves were obtained using the SPICE model in Table~\ref{tbl:memR} (Appendix B).}
   \label{fig:memR}
  \end{center}
\end{figure}

Fig.~\ref{fig:memR} shows current-voltage curves for the memristive element. To obtain these curves, we used the SPICE model from Table~\ref{tbl:memR}. According to Fig.~\ref{fig:memR},
the memristance decreases at $V_M>0$ and increases at $V_M<0$.
 The frequency dependence of current-voltage curves in Fig.~\ref{fig:memR} is typical for memristive systems~\cite{chua76a,09_memelements}.

 We have found that the polarity of memristive elements in Fig.~\ref{fig:1} is incorrect. To understand this, one can evaluate the total terminal current in a gate terminal assuming a logically-consistent state.  As we have mentioned above, in this case, the total terminal current must be zero (see the second paragraph in Sec.~\ref{sec2_1}). For instance, a simple calculation shows that the current in the terminal 1 of AND at $V_1=V_2=V_o=-1$~V is $I_1=2/M_1-2/R_{off}$ and the voltage across $M_1$ in Fig.~\ref{fig:1} circuit is $+2$~V. As positive voltages drive $M_1$ into $R_{on}$, $I_1=2/R_{on}-2/R_{off}>0$. To satisfy $I_1=0$, the polarity of $M_1$ must be reversed.

As all variables $s_j$ are described by the same Eq.~(\ref{eq:6}), we represent these variables by a single variable $s\equiv s_j$ for all $j$. In SPICE, we have defined an s-block as the component that integrates Eq.~(\ref{eq:6}) (see the second model in Table~\ref{tbl:VCDCG}). The s-block has $8$ voltage inputs for the voltage signals encoding $i_{DCD,j}$-s and a single output, which is $s$. The output of s-block must be connected to all VCDCGs (to terminals 4). The inputs of s-block are taken from terminals 3 of VCDCGs. The SPICE model of s-block in Table~\ref{tbl:VCDCG} can be directly used with up to $8$  VCDCGs in the circuit and can be straightforwardly expanded to a larger number of VCDCGs.

 \begin{figure}
 \begin{center}
  \includegraphics[width=0.7\columnwidth,keepaspectratio]{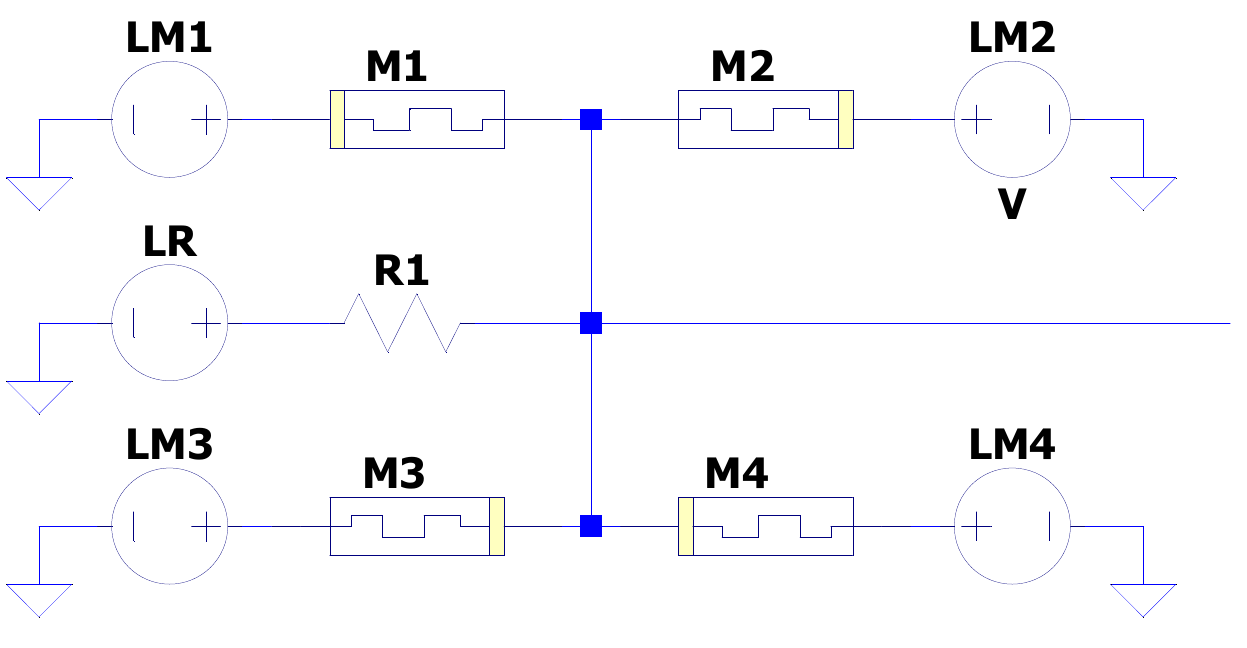}
   \fcaption{Correct schematics of the dynamic correction module.}
   \label{fig:DCM}
  \end{center}
\end{figure}

\begin{figure}
 \begin{center}
(a)\hspace{0.5cm}\includegraphics[width=0.6\columnwidth,keepaspectratio]{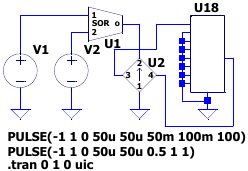}   
  \\     \vspace{0.3cm}
  (b)\includegraphics[width=0.8\columnwidth,keepaspectratio]{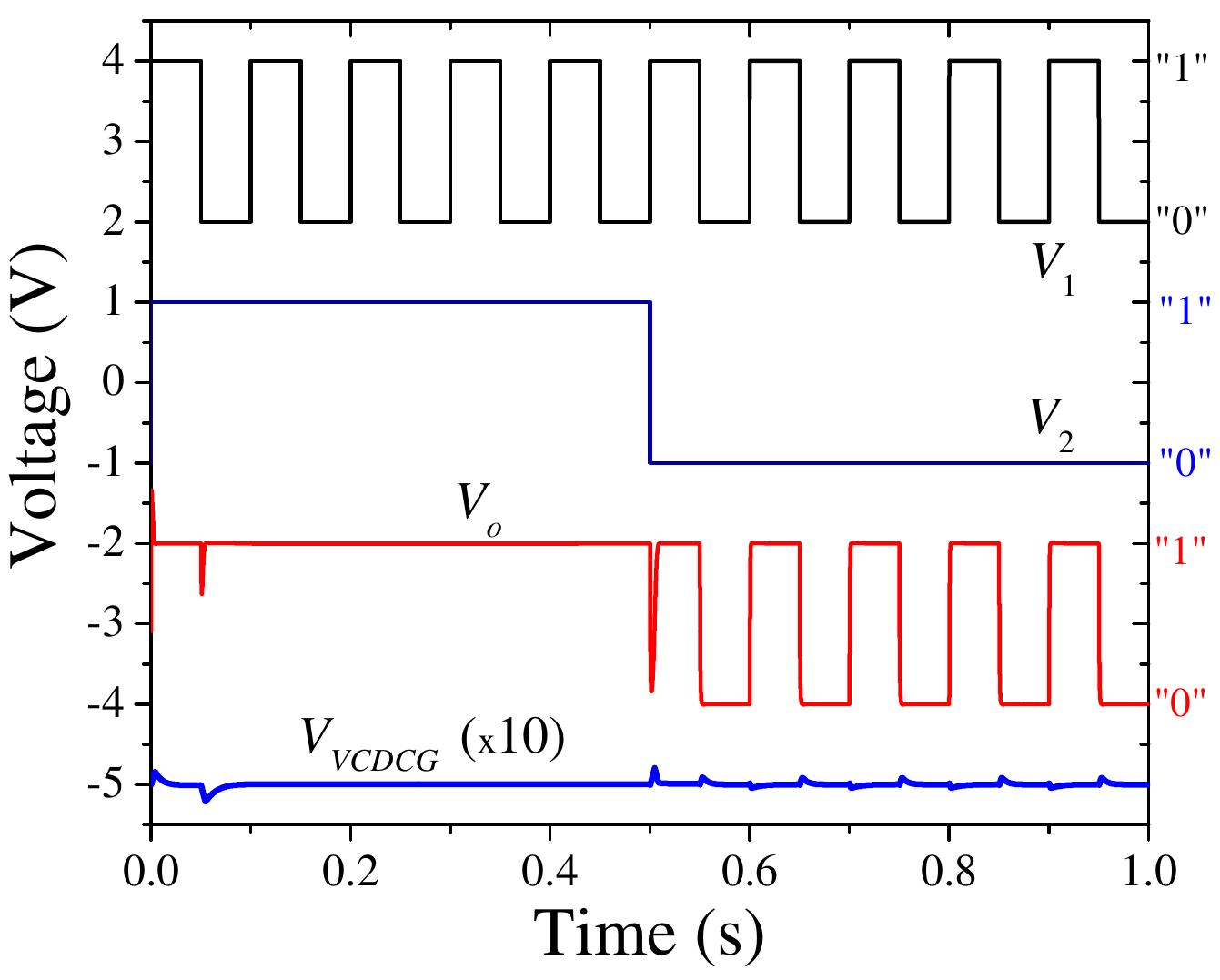}
   \fcaption{Direct operation of self-organizing OR. (a) Circuit used in simulations. Here, V1 and V2 are the pulsed voltage sources, U1 is the self-organizing OR, U2 is the VCDCG, and U18 is the s-block.   
   (b) Voltage transient signals at the terminals of self-organizing OR and voltage output of VCDCG (terminal 3 of U2).
   The curves were displaced for clarity.
   }
   \label{fig:directSOR}
  \end{center}
\end{figure}  

\rengSubsection{Important implementation notes} \label{sec:3_2}

\begin{enumerate}
  \item The correct version of Design I self-organizing gates involves two resistors (Fig.~\ref{fig:1}, left).
  In~\cite{MemComputingbook}, it is shown without resistors. In~\cite{di2018self}, it is presented with memristors instead of resistors.
  \item The polarity of memristors in Fig.~\ref{fig:1} has been reversed, see Fig.~\ref{fig:DCM}.
  \item We emphasize that unlike Refs.~\cite{Traversa17a,di2018self}, we use the minus sign in front of $k_i$ in Eq.~(\ref{eq:8}).
  \item The values $k_i=10^{-7}$ and $k_s=10^{-7}$ from~\cite{Traversa17a,di2018self} are too small. To enable the correct operation of the reset feature in  VCDCGs, we use $k_i=k_s=$2E3.   
  \item Memristive elements subjected to zero voltage and associated voltage sources have not been included in the models of AND and OR (e.g., $L_{M_2}$ and $L_{M_4}$ and associated memristive elements in the DCM of terminal 1 of AND).
  
\begin{figure*}
 \begin{center}
(a)\hspace{0.4cm}\includegraphics[width=0.35\textwidth,keepaspectratio]{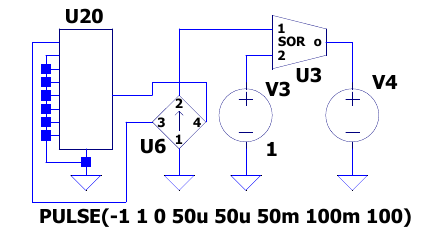} 
(b)\hspace{0.4cm}\includegraphics[width=0.32\textwidth,keepaspectratio]{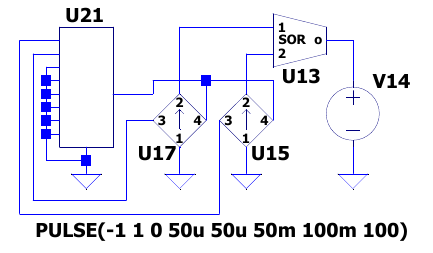} \\
(c)\includegraphics[width=0.4\textwidth,keepaspectratio]{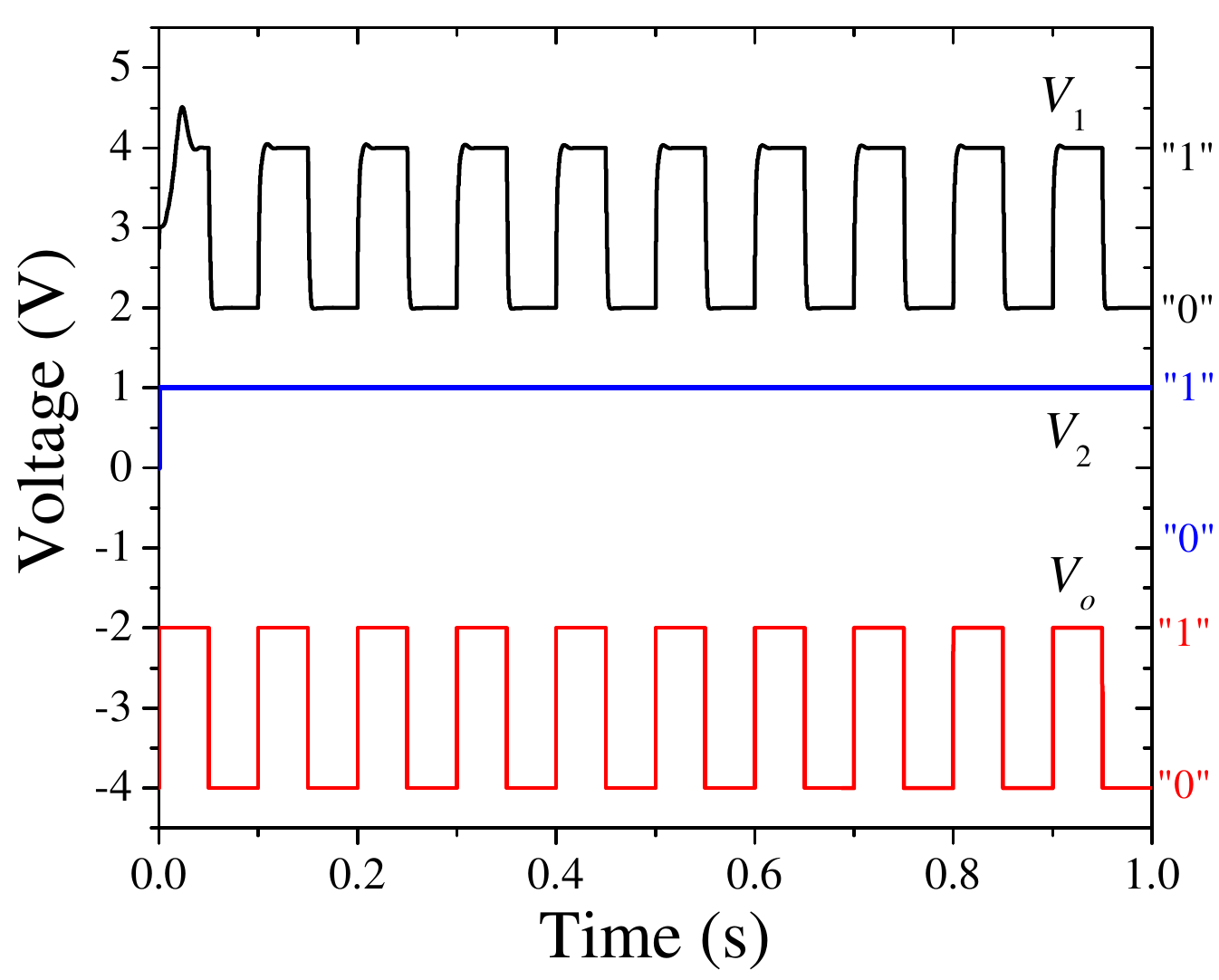}
(d)\includegraphics[width=0.4\textwidth,keepaspectratio]{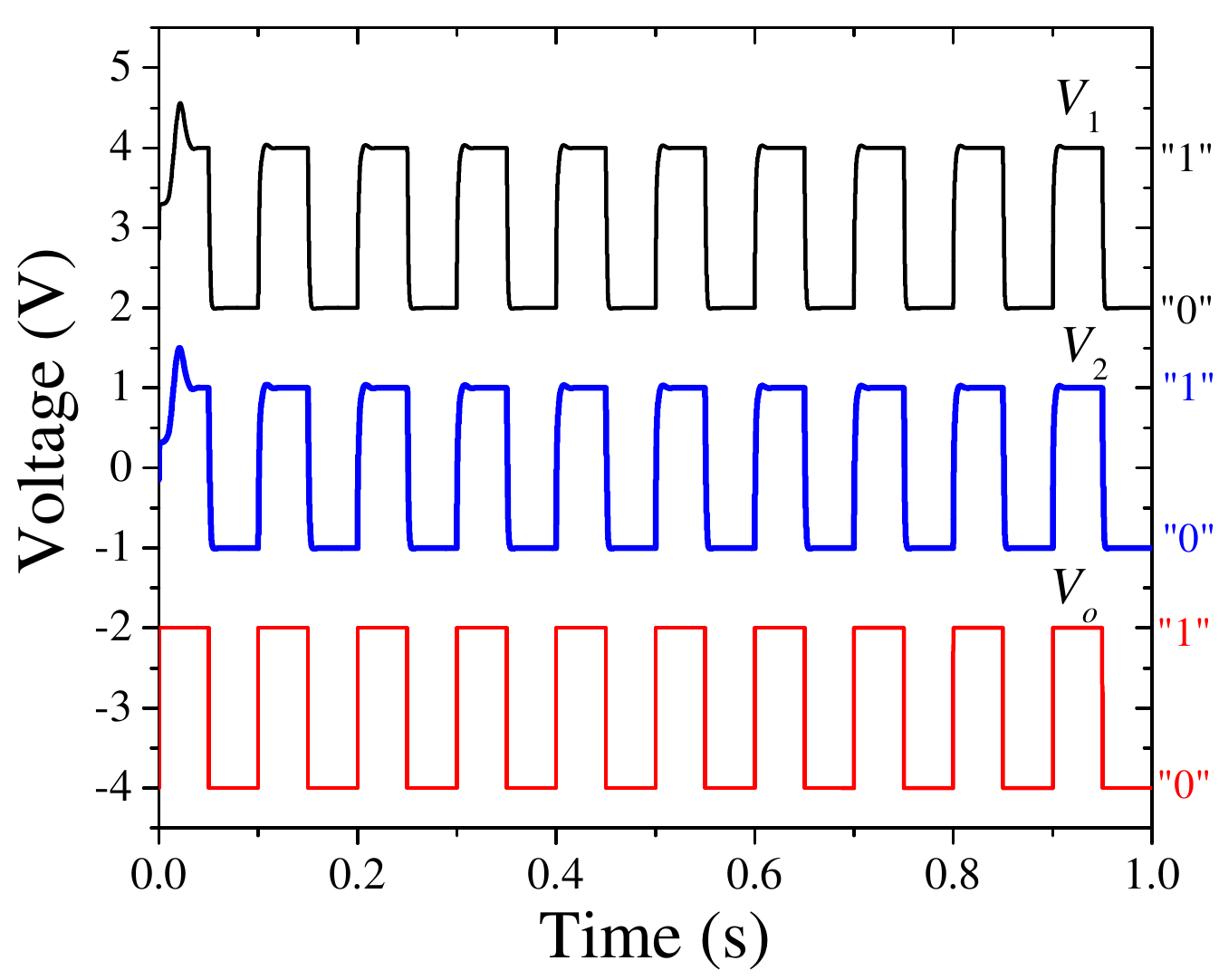}
   \fcaption{Reverse operation of self-organizing OR. (a), (b) Circuits used in simulations. Here, V4 and V14 are pulsed voltage sources, and V1 is the $1$~V constant voltage source. (c) 
 Example of voltage transient signals at the terminals of self-organizing OR in (a). (d) Example of
   voltage transient signals at the terminals of self-organizing OR in (b).}
   \label{fig:reverseSOR}
  \end{center}
\end{figure*}

  \item All $s_j$-s are implemented using a single variable $s$.
  \item To suppress high-voltage spikes~\footnote{In the circuit based on original parameters, spikes can be of quite extreme magnitude (e.g., several hundred thousand volts). These spikes have been associated with instantons, 
  see~\cite{MemComputingbook}.} and improve the convergence, we have increased $R_{on}$ to $0.05$, decreased $i_{max}$ to 10, and added capacitor across each VCDCG (C1 in Table~\ref{tbl:VCDCG}).
  \item To enable the random initial states of memristive elements, the option "Use the clock to reseed the MC generator" must be checked in  LTSpice XVII. 
The transient analysis was performed using the option {\it uic}.
  \item For reproducibility of our results, the initial states of memristive elements are chosen from a flat random distribution between $0.18$ and $0.22$. All other initial values are selected deterministically.
\end{enumerate}

\begin{figure*}
 \begin{center}
(a)\includegraphics[width=0.9\textwidth,keepaspectratio]{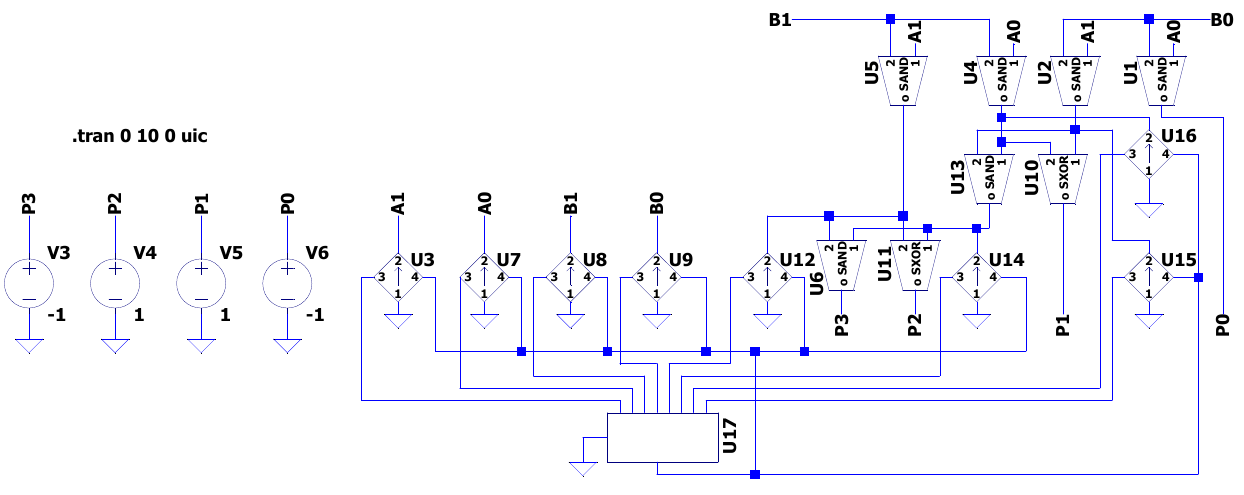}\\
(b)\hspace{0.4cm}\includegraphics[width=0.25\textwidth,keepaspectratio]{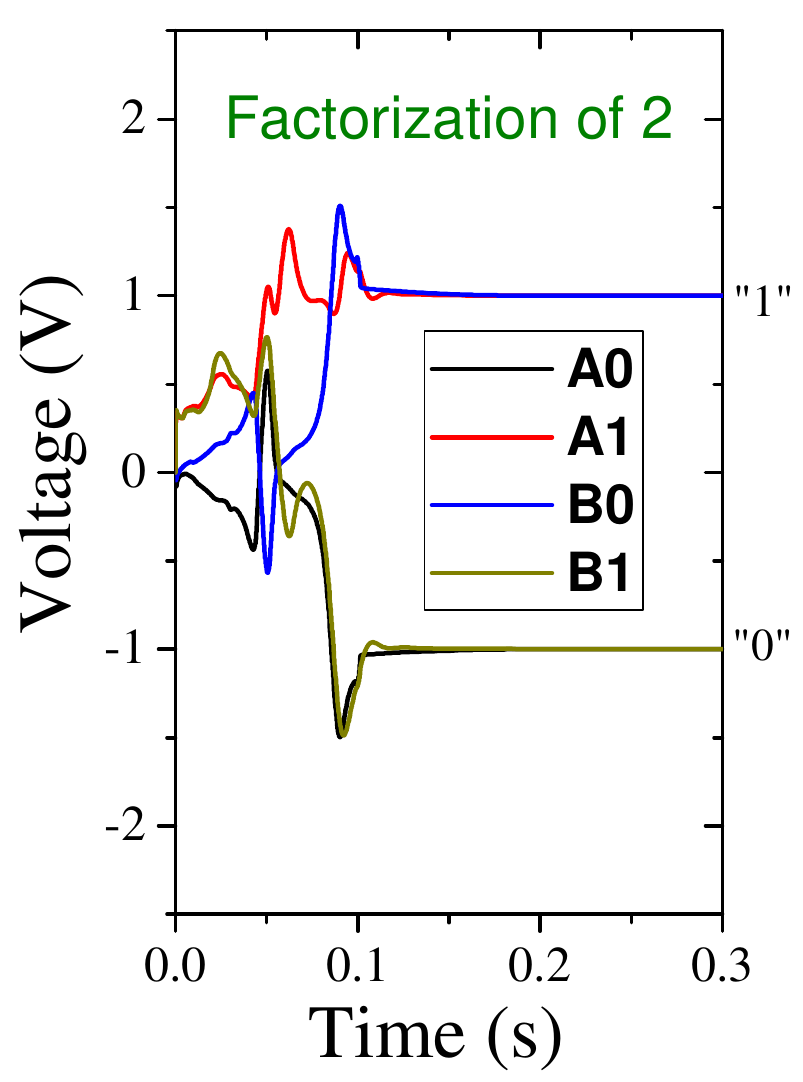} (c)\hspace{0.4cm}\includegraphics[width=0.25\textwidth,keepaspectratio]{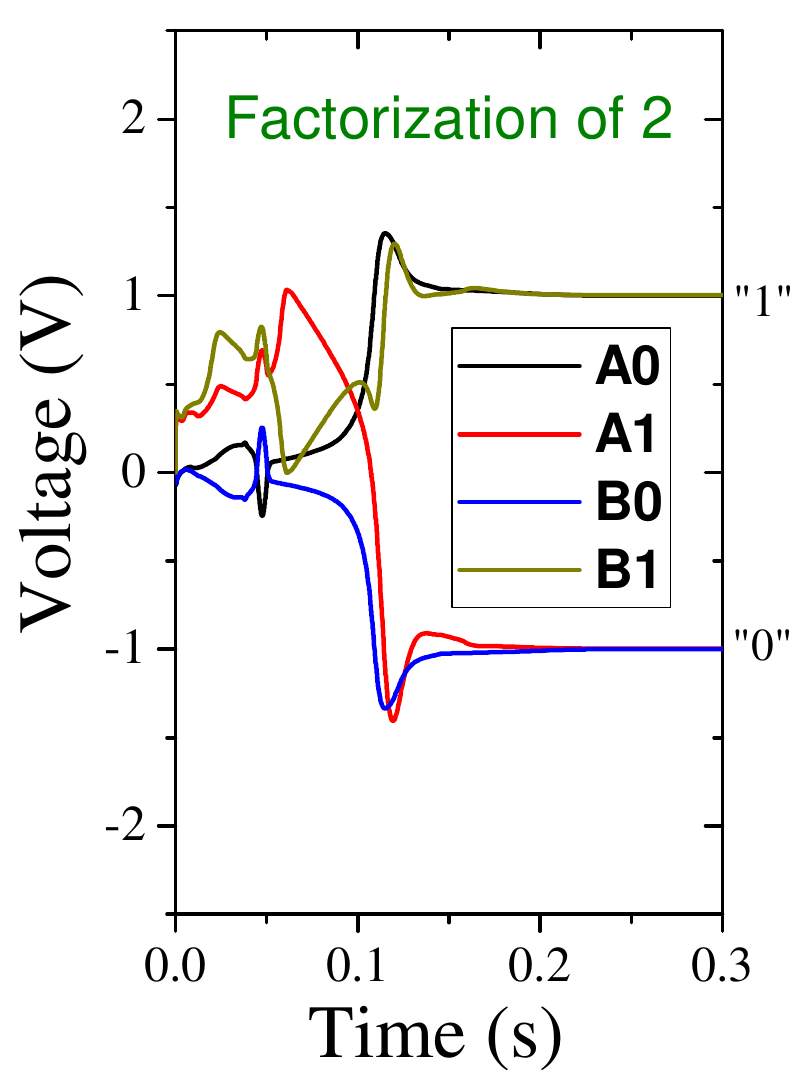} 
(d) \hspace{0.4cm}\includegraphics[width=0.25\textwidth,keepaspectratio]{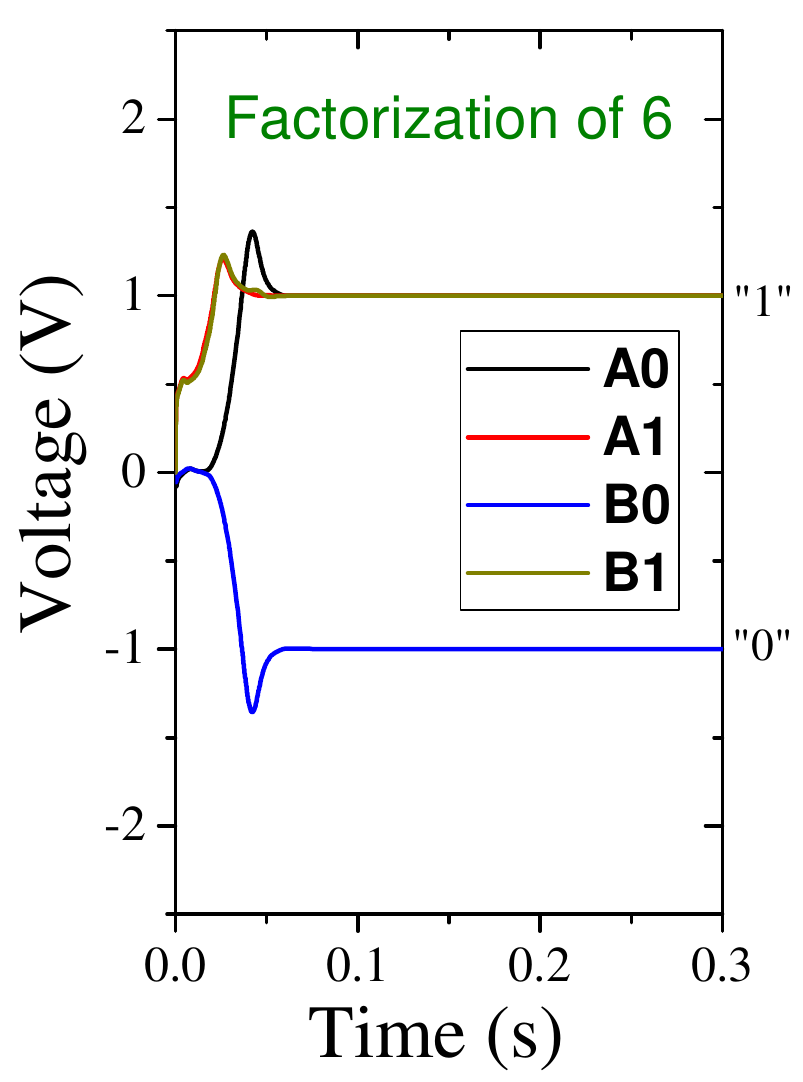} \\
(e)\includegraphics[width=0.4\textwidth,keepaspectratio]{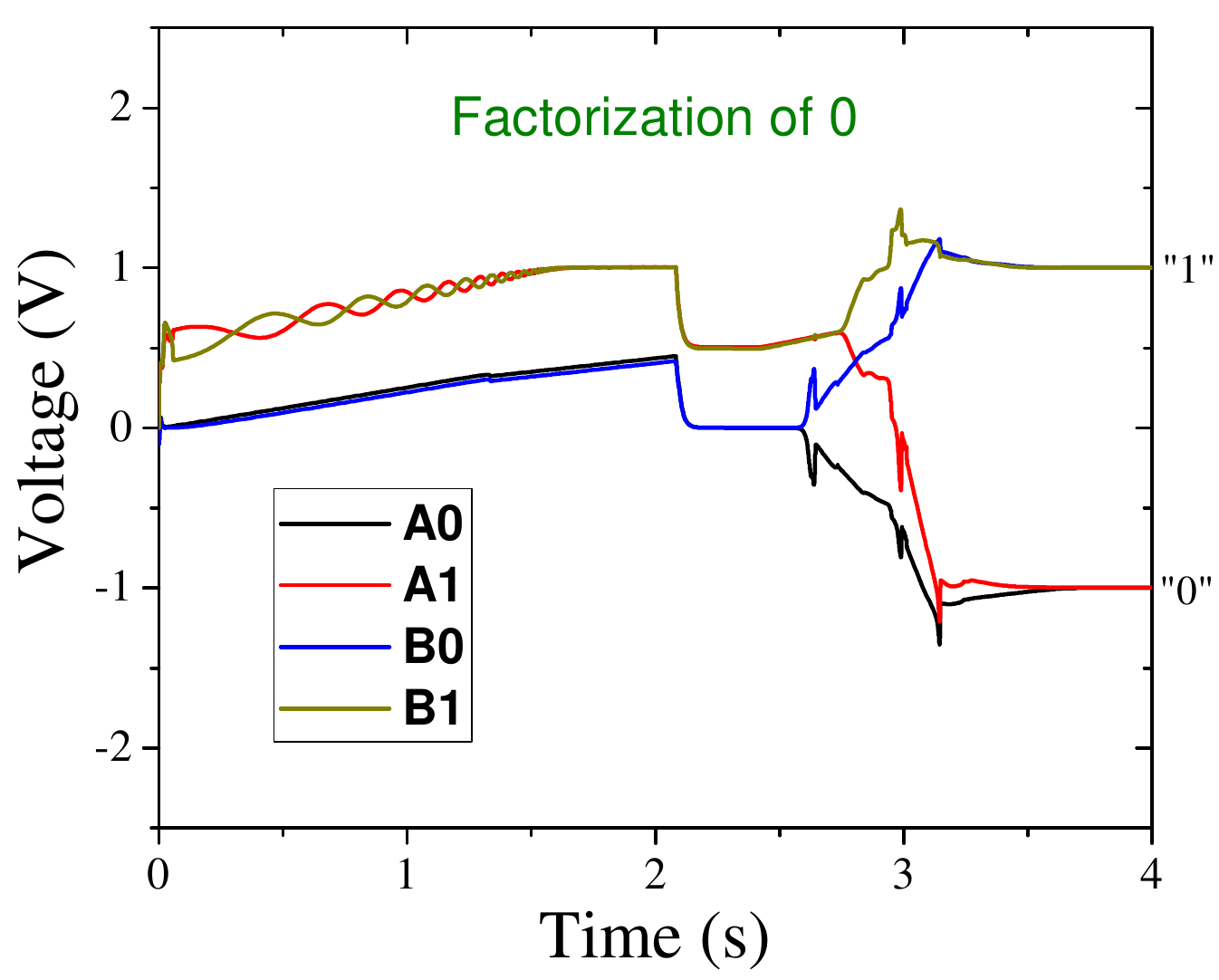}
(f)\includegraphics[width=0.4\textwidth,keepaspectratio]{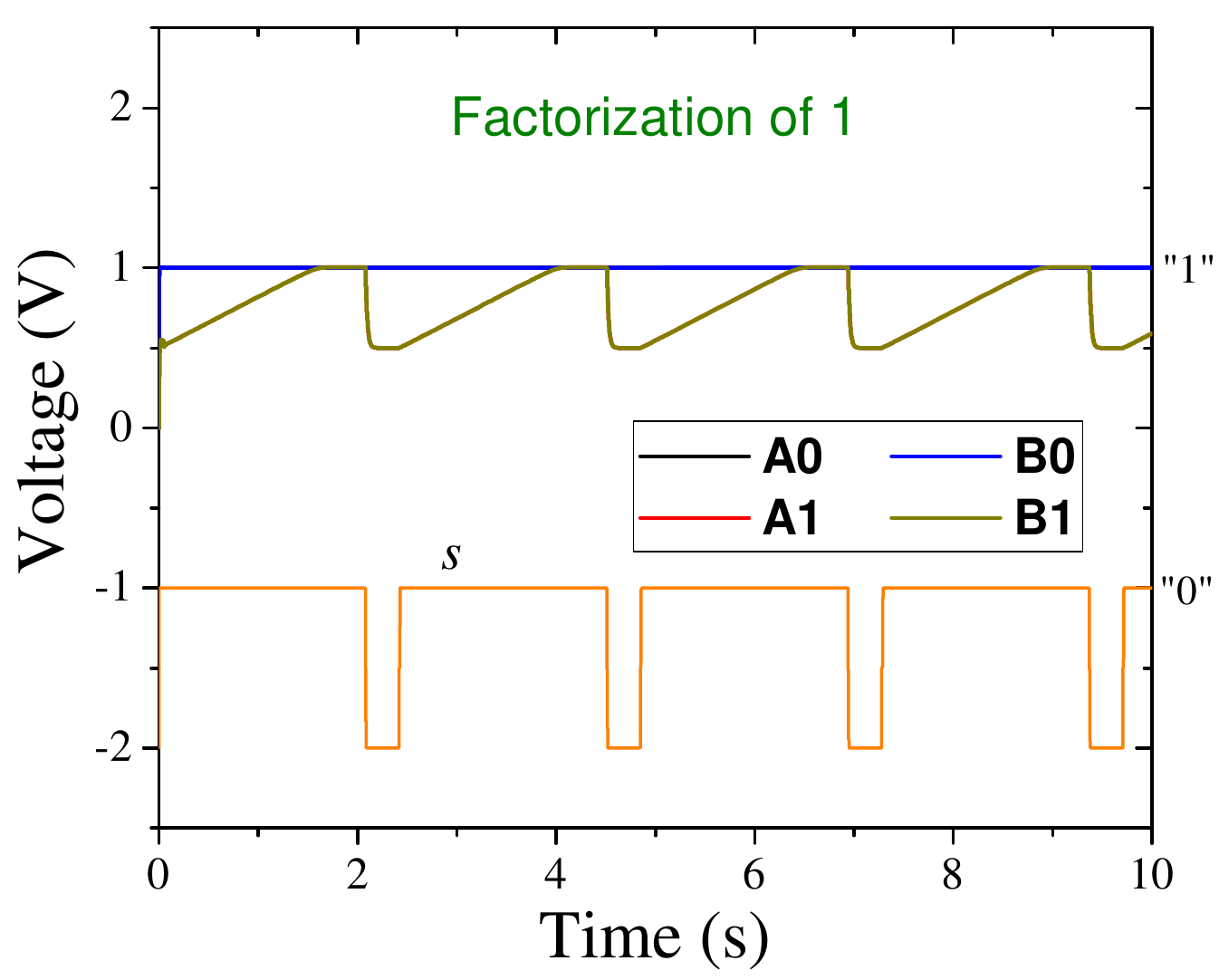}
   \fcaption{Solving integer factorization problem with a self-organizing 2-bit by 2-bit multiplier. (a)  Circuit used in simulations.
   Here, the number to factorize is represented by (P3,P2,P1,P0), and the factors -- by (A1,A0) and (B1,B0). (b)-(f) Examples of the transient dynamics of the self-organizing 2-bit by 2-bit multiplier. Curve $s$ in (f) was shifted down by 2~V for clarity.}
   \label{fig:s2b2}
  \end{center}
\end{figure*}

\rengSection{Simulation examples}  \label{sec:4}

\rengSubsection{Single gates}

This subsection exemplifies the behavior of self-organizing gates using the OR gate as an example.
First, we consider the traditional (direct) operation, wherein the voltage signals are applied to the 
traditional inputs of the gate. Second, we explore the reverse operation (not available with the usual OR).

Fig.~\ref{fig:directSOR}(a) shows the simulated circuit  for the direct operation of self-organizing OR. 
In this circuit, the input signals are applied using pulsed voltage sources connected to terminals 1 and 2 of U1. The output terminal of U1 (not connected to any voltage source) is driven by a VCDCG (see the circuit transformation rules in Sec.~\ref{sec:2}).  Fig.~\ref{fig:directSOR}(b)  shows the gate response.
Clearly, except of short transients, the gate reproduces the truth table of OR.
In Fig.~\ref{fig:directSOR}(b), the curve $V_{VCDCG}$ represents the current into VCDCG U2 (without accounting for C1's current). We note that this current fluctuates at about zero.

Next, consider the reverse operation of self-organizing OR. Figs.~\ref{fig:reverseSOR}(a) and (b) show two slightly different circuits used in our simulations. 
The difference is that in Fig.~\ref{fig:reverseSOR}(a) we use two terminals (terminals 2 and $o$) as inputs and one terminal (1) as output, while in Fig.~\ref{fig:reverseSOR}(b)
there is a single input terminal ($o$) and two output terminals (1 and 2).
 As before, we use VCDCGs to ensure the binary states at the output terminals.

Fig.~\ref{fig:reverseSOR}(c) shows the gate voltages for the circuit in Fig.~\ref{fig:reverseSOR}(a).
Note that after a relatively short transient, the voltage at the first terminal, $V_1(t)$, repeats the applied voltage, $V_o(t)$. When $V_o=1$~V, the truth table of OR is satisfied. In the opposite case, the gate shows reasonable behavior as it chooses $V_1=-1$~V over $V_1=1$~V deterministically. 
In some of the other runs, instead of the pulsed voltage at terminal 1, we observed $V_1(t)\approx -1$~V (after a transient interval). In this case, again, the truth table of OR is satisfied whenever $V_o(t)=1$~V.

Finally, consider the response of self-organizing OR in  Fig.~\ref{fig:reverseSOR}(b). 
Using a flat random distribution of initial states of memristive elements from 0 to 1, in each run we observed one of the following general responses: (i) $V_1(t)=V_2(t)=V_o(t)$ (as in Fig.~\ref{fig:reverseSOR}(d)),  $V_1(t)=-1$~V, $V_2(t)=V_o(t)$, or (iii) $V_1(t)=V_o(t)$, $V_2(t)=-1$~V. Clearly, all these cases are consistent with the truth table of OR.

Overall, we conclude that the self-organizing OR reproduces correctly the truth table of OR (whenever possible) no matter the role of each terminal as input or output. The response may be different in different runs (but always correct). We have verified that the same is true for self-organizing AND and XOR.

\rengSubsection{Circuits of self-organizing gates}

As an example of a circuit of self-organizing gates, consider a self-organizing 2-bit by 2-bit multiplier. Its schematics is presented in
Fig.~\ref{fig:s2b2}(a). The self-organizing multiplier involves eight self-organizing gates and eight voltage-controlled differential current generators. This circuit was designed based on the conventional  2-bit by 2-bit binary multiplier using the circuit transformation rules outlined in  Sec.~\ref{sec:2}.

The circuit in Fig.~\ref{fig:s2b2}(a) uses four constant voltage sources V3-V6 to encode the number to factorize (it is 6 in Fig.~\ref{fig:s2b2}(a)). We emphasize that the signals P0-P3 serve as the input. The output signals are A0, A1, B0, and B1, which are the bits of two factors (A0 and B0 are the least significant bits). These factors are found through the deterministic dynamics of the self-organizing multiplier.

Examples of circuit dynamics are demonstrated in Fig.~\ref{fig:s2b2}(b)-(f). In particular, Fig.~\ref{fig:s2b2}(b) and (c) show that the result may be different in different runs. Specifically, these plots indicate that number 2 can be presented as $2\cdot 1$ or $1\cdot 2$. This ability to identify different solutions is related to the random choice of initial states of memristive elements. Two distinct solutions are observed when two sets of initial states belong to different basins of attraction.

Figs.~\ref{fig:s2b2}(d) and (e) indicate that the factorization of some numbers can be more difficult than others (using some close initial conditions). In our simulations of the self-organizing multiplier, the most difficult was the factorization of 1. In this case, most frequently, we have observed the transition to a limit cycle behavior  as the one in Fig.~\ref{fig:s2b2}(f) and, quite occasionally, the correct solution to the problem ($1=1\cdot 1$).  

We have verified that the existence of the limit cycle (Fig.~\ref{fig:s2b2}(f)) is not related to certain modifications to the parameters that we made. In particular, the limit cycle was observed in the circuit without C1 (in the model of VCDCG), and with prior values of $R_{on}$, $q$, and $i_{max}$ (from Table~\ref{tbl:params}). In a longer simulation, it was observed that the limit cycle continues up to $100$~s. To ensure that the limit cycle is not a numerical artifact, we have performed some additional simulations. It was observed that the use of other numerical integration methods in LTspice (trapezoid and modified trap in addition to Gear)~\footnote{It is known that the basin of attraction is influenced by the discretization~\cite{zhang2021directed}.}, variation of tolerances, the addition of noise, and the use of PSPICE result in the same limit cycle behavior.

The existence of the limit cycle in the numerical dynamics seems to contradict the statement "if the Boolean problem the DMMs are designed to solve has a solution, the system will always find it, irrespective of the initial conditions" in Ref.~\cite{no-chaosb} (see also~\cite{Traversa17a}) for the continuous dynamics. Currently, the exact reason for this is not known and its determination is beyond the scope of this work.

\rengSection{Conclusion} \label{sec:5}

Having identified and corrected some inconsistencies in the prior literature~\cite{Traversa17a,di2018self,MemComputingbook} (see items 1-4 in Sec.~\ref{sec:3_2}), we have formulated SPICE models of the Design I self-organizing logic gates. The operation of individual self-organizing gates and small circuits thereof has been demonstrated. 

We emphasize that in future studies of these gates, special attention should be paid among others to:
\begin{itemize}
    \item Polarity of memristive devices in DCMs.
    \item Use of resistors in the schematics of the universal gate.
    \item Sign of $k_i$ term in the function $f_s\left( \mathbf{i}_{DCG},s_j\right)$.
    \item Values for parameters $k_i$ and $k_s$.
\end{itemize}

In summary, self-organizing logic gates are an interesting generalization of the traditional Boolean logic gates.  Being designed primarily to solve complex optimization problems, these gates may also be useful for other applications. The SPICE models reported in this paper offer an easy and pretty reliable way to explore self-organization in memcomputing circuits. 

%\section*{Declaration of Competing Interest}

%The authors declare that they have no known competing financial interests or personal relationships that could have appeared to influence the work reported in this paper.




\rengAck

The author acknowledges the support from the National Science Foundation grant number ECCS-2229880. He is thankful to M. Di Ventra, D. C. Nguyen, and Y.-H. Zhang for numerous discussions on various aspects of memcomputing.

\bibliographystyle{elsarticle-num}
\bibliography{Xbibli}

\end{multicols}

\newpage

\renewcommand\thetable{A.\arabic{table}}

\rengAppendix{Parameters} 

\vspace{1cm}

\begin{table}
\begin{tabular}{l|cccc|cccc|cccc}
\hline
\hline
 & \multicolumn{4}{c|}{Terminal 1} & \multicolumn{4}{c|}{Terminal 2} & \multicolumn{4}{c}{Out Terminal} \\
 & $a_1$ & $a_2$ & $a_0$ & $dc$ & $a_1$ & $a_2$ & $a_0$ & $dc$ &$a_1$ & $a_2$ & $a_0$ & $dc$\\
 \hline
\multicolumn{13}{l}{SO AND} \\
$L_{M_1}$ & 0 & -1 & 1 & $v_c$ & -1 & 0 & 1 & $v_c$ & 1 & 0 & 0 & 0 \\
$L_{M_2}$ & 1 & 0 & 0 & 0 & 0 & 1 & 0 & 0 & 0 & 1 & 0 & 0 \\
$L_{M_3}$ & 0 & 0 & 1 & 0 & 0 & 0 & 1 & 0 & 0 & 0 & 1 & 0 \\
$L_{M_4}$ & 1 & 0 & 0 & 0 & 0 & 1 & 0 & 0 & 2 & 2 & -1 & $-2v_c$ \\
$L_{R}$ & 4 & 1 & -3 & $-v_c$ & 1 & 4 & -3 & $-v_c$ & -4 & -4 & 7 & $2v_c$ \\
\multicolumn{13}{l}{SO OR} \\
$L_{M_1}$ & 0 & 0 & 1 & 0 & 0 & 0 & 1 & 0 & 0 & 0 & 1 & 0 \\
$L_{M_2}$ & 1 & 0 & 0 & 0 & 0 & 1 & 0 & 0 & 2 & 2 & -1 & $2v_c$ \\
$L_{M_3}$ & 0 & -1 & 1 & $-v_c$ & -1 & 0 & 1 & $-v_c$ & 1 & 0 & 0 & 0 \\
$L_{M_4}$ & 1 & 0 & 0 & 0 & 0 & 1 & 0 & 0 & 0 & 1 & 0 & 0 \\
$L_{R}$ & 4 & 1 & -3 & $v_c$ & 1 & 4 & -3 & $v_c$ & -4 & -4 & 7 & $-2v_c$ \\
\multicolumn{13}{l}{SO XOR} \\
$L_{M_1}$ & 0 & -1 & -1 & $v_c$ & -1 & 0 & -1 & $v_c$ & -1 & -1 & 0 & $v_c$ \\
$L_{M_2}$ & 0 & 1 & 1 & $v_c$ & 1 & 0 & 1 & $v_c$ & 1 & 1 & 0 & $v_c$ \\
$L_{M_3}$ & 0 & -1 & 1 & $-v_c$ & -1 & 0 & 1 & $-v_c$ & -1 & 1 & 0 & $-v_c$ \\
$L_{M_4}$ & 0 & 1 & -1 & $-v_c$ & 1 & 0 & -1 & $-v_c$ & 1 & -1 & 0 & $-v_c$ \\
$L_{R}$ & 6 & 0 & -1 & 0 & 0 & 6 & -1 & 0 & -1 & -1 & 7 & 0 \\ \hline
\end{tabular}
\caption{\label{tbl:1} Parameters of voltage-controlled voltage generators for AND, OR, and XOR gates.}
\end{table}

\vspace{1 cm}

\begin{table}
\begin{tabular}{llllll}
\hline \hline
Parameter & Value & Parameter & Value & Parameter & Value \\
\hline
$R_{on}$ & $10^{-2}$ & $R_{off}$ & 1 & $v_c$ & 1 \\
$\alpha$ & 60 & $C$ & $10^{-9}$ & $k$ & $\infty$ \\
$V_t$ & 0 & $\gamma$ & 60 & $q$ & 10 \\
$m_0$ & -400 & $m_1$ & 400 & $i_{min}$ & $10^{-8}$ \\
$i_{max}$ & 20 & $k_i$ & $10^{-7}$ & $k_s$ & $10^{-7}$ \\
$\delta_s$ & 0 & $\delta_i$ & 0 &  & \\
\hline
\hline
\end{tabular}
\caption{\label{tbl:params} Parameters of numerical simulations used in Ref.~\cite{Traversa17a}.}
\end{table}

\newpage

\rengAppendix{SPICE models \label{app:models}}

\setcounter{table}{0}
\renewcommand\thetable{B.\arabic{table}}

\vspace{1cm}

\begin{table}
\begin{tabular}{l}
**** Self-organizing AND *********************************************\\
**** Code for LTspice; tested with LTspice XVII***********************\\
**********************************************************************\\
.subckt SAND 1 2 3\\
.param res=1 vc=1\\
* DCM1\\
EM11 11 0 value=\{-V(2)+V(3)+vc\}\\
Xmem11 1 11 memR\\
EM14 14 0 value=\{V(3)\}\\
Xmem14 14 1 memR\\
EM13 13 0 value=\{4*V(1)+1*V(2)-3*V(3)-vc\}\\
R11 13 1 \{res\}\\
* DCM2\\
EM21 21 0 value=\{-V(1)+V(3)+vc\}\\
Xmem21 2 21 memR\\
EM24 24 0 value=\{V(3)\}\\
Xmem24 24 2 memR\\
EM23 23 0 value=\{V(1)+4*V(2)-3*V(3)-vc\}\\
R21 23 2 \{res\}\\
* DCM3\\
EM31 31 0 value=\{V(1)\}\\
Xmem31 3 31 memR\\
EM32 32 0 value=\{V(2)\}\\
Xmem32 3 32 memR\\
EM35 35 0 value=\{2*V(1)+2*V(2)-1*V(3)-2*vc\}\\
Xmem35 35 3 memR\\
EM33 33 0 value=\{-4*V(1)-4*V(2)+7*V(3)+2*vc\}\\
R31 33 3 \{res\}\\
* resistors\\
R1o 1 3 \{res\}\\
R2o 2 3 \{res\}\\
.ends SAND\\
\end{tabular}
\caption{\label{tbl:SAND} LTspice code for the self-organizing AND.}
\end{table}

\vspace{1cm}
\begin{table}
\begin{tabular}{l}
**** Self-organizing XOR *********************************************\\
**** Code for LTspice; tested with LTspice XVII***********************\\
**********************************************************************\\
.subckt SXOR 1 2 3\\
.param res=1 vc=1\\
* DCM1\\
EM11 11 0 value=\{-V(2)-V(3)+vc\}\\
Xmem11 1 11 memR\\
EM12 12 0 value=\{V(2)+V(3)+vc\}\\
Xmem12 1 12 memR\\
EM14 14 0 value=\{-V(2)+V(3)-vc\}\\
Xmem14 14 1 memR\\
EM15 15 0 value=\{V(2)-V(3)-vc\}\\
Xmem15 15 1 memR\\
EM13 13 0 value=\{6*V(1)-V(3)\}\\
R11 13 1 \{res\}\\
* DCM2\\
EM21 21 0 value=\{-V(1)-V(3)+vc\}\\
Xmem21 2 21 memR\\
EM22 22 0 value=\{V(1)+V(3)+vc\}\\
Xmem22 2 22 memR\\
EM24 24 0 value=\{-V(1)+V(3)-vc\}\\
Xmem24 24 2 memR\\
EM25 25 0 value=\{V(1)-V(3)-vc\}\\
Xmem25 25 2 memR\\
EM23 23 0 value=\{6*V(2)-V(3)\}\\
R21 23 2 \{res\}\\
* DCM3\\
EM31 31 0 value=\{-V(1)-V(2)+vc\}\\
Xmem31 3 31 memR\\
EM32 32 0 value=\{V(1)+V(2)+vc\}\\
Xmem32 3 32 memR\\
EM34 34 0 value=\{-V(1)+V(2)-vc\}\\
Xmem34 34 3 memR\\
EM35 35 0 value=\{V(1)-V(2)-vc\}\\
Xmem35 35 3 memR\\
EM33 33 0 value=\{-V(1)-V(2)+7*V(3)\}\\
R31 33 3 \{res\}\\
* resistors\\
R1o 1 3 \{res\}\\
R2o 2 3 \{res\}\\
.ends SXOR\\
\end{tabular}
\caption{\label{tbl:SXOR} LTspice code for the self-organizing XOR.}
\end{table}

\vspace{1cm}
\begin{table}
\begin{tabular}{l}
**** Self-organizing OR **********************************************\\
**** Code for LTspice; tested with LTspice XVII***********************\\
**********************************************************************\\
.subckt SOR 1 2 3\\
.param res=1 vc=1\\
* DCM1\\
EM11 11 0 value=\{V(3)\}\\
Xmem11 1 11 memR\\
EM14 14 0 value=\{-V(2)+V(3)-vc\}\\
Xmem14 14 1 memR\\
EM13 13 0 value=\{4*V(1)+1*V(2)-3*V(3)+vc\}\\
R11 13 1 \{res\}\\
* DCM2\\
EM21 21 0 value=\{V(3)\}\\
Xmem21 2 21 memR\\
EM24 24 0 value=\{-V(1)+V(3)-vc\}\\
Xmem24 24 2 memR\\
EM23 23 0 value=\{V(1)+4*V(2)-3*V(3)+vc\}\\
R21 23 2 \{res\}\\
* DCM3\\
EM32 32 0 value=\{2*V(1)+2*V(2)-V(3)+2*vc\}\\
Xmem32 3 32 memR\\
EM34 34 0 value=\{V(1)\}\\
Xmem34 34 3 memR\\
EM35 35 0 value=\{V(2)\}\\
Xmem35 35 3 memR\\
EM33 33 0 value=\{-4*V(1)-4*V(2)+7*V(3)-2*vc\}\\
R31 33 3 \{res\}\\
* resistors\\
R1o 1 3 \{res\}\\
R2o 2 3 \{res\}\\
.ends SOR\\
\end{tabular}
\caption{\label{tbl:SOR} LTspice code for the self-organizing OR.}
\end{table}

\vspace{1cm}
\begin{table}
\begin{tabular}{l}
**** Memristive element **********************************************\\
**** Code for LTspice; tested with LTspice XVII***********************\\
**********************************************************************\\
.subckt memR plus minus PARAMS:\\
+ Ron=0.05 Roff=1 alpha=60 C=1E-9\\
*model of memristive port\\
Gpm plus minus value=\{V(plus,minus)/(Ron+(Roff-Ron)*V(x))\}\\
C1 plus minus \{C\} IC=\{0\}\\
*integrator model\\
Gx 0 x value=\{-alpha*h(V(x),V(plus,minus))*V(plus,minus)/(Ron+(Roff-Ron)*V(x))\}\\
Cx x 0 1 IC=\{mc(0.2,0.1)\} ; randomized initial condition\\
Raux x 0 100meg\\
* functions\\
.func h(x,vm)=\{u(x)*u(vm)+u(1-x)*u(-vm)\}\\
.ends memR\\
\end{tabular}
\caption{\label{tbl:memR} LTspice code for the memristive element.}
\end{table} 

\vspace{1cm}
\begin{table}
\begin{tabular}{l}
**** Voltage-Controlled Differential Current Generator ***************\\
**** Code for LTspice; tested with LTspice XVII***********************\\
**********************************************************************\\
.subckt VCDCG 1 2 3 4\\
* 1: negative terminal (ground); 2: positive terminal;\\
* 3: current output signal (voltage); 4: input from the s-circuit\\
.param m0=-400 m1=400 q=5 gamma=60\\
* model of VCDCG ports\\
G1 2 1 value=\{V(x)\}\\
E1 3 1 value=\{V(x)\}\\
R1 4 1 100meg\\
C1 2 1 1E-3 IC=\{0\}\\
* integrator model\\
Gx 0 x value = \{u(V(4,1)-0.5)*fdcg(V(2,1))-gamma*u(0.5-V(4,1))*V(x)\}\\
Cx x 0 1 IC=\{0\}\\
Raux x 0 100meg\\
* functions\\
.param m0b=\{m0*pi/(2*q)\} m1b=\{m1*pi/(2*q)\}\\
.func fdcg(x)=\{q*(atan(m1b*(x+1))+atan(m0b*x)+atan(m1b*(x-1)))*2/pi\}\\
.ends VCDCG\\
\\
\\
**** 8-input s-block *************************************************\\
**** Code for LTspice; tested with LTspice XVII***********************\\
**********************************************************************\\
.subckt SCOMB8 1 2 3 4 5 6 7 8 9 10\\
* 1: negative terminal (ground); 2: positive terminal (output); 3-10: inputs\\
.param imin=1E-8 imax=10 ks=2E3 ki=2E3\\
* model\\
E1 2 1 value=\{V(s)\}\\
R1 1 3 100meg\\
R2 1 4 100meg\\
R3 1 5 100meg\\
R4 1 6 100meg\\
R5 1 7 100meg\\
R6 1 8 100meg\\
R7 1 9 100meg\\
R8 1 10 100meg\\
*integrator model\\
Gs 0 s value =\{-ks*V(s)*(V(s)-1)*(2*V(s)-1)-\\
+ki*(1-u(imin-abs(V(3,1)))*u(imin-abs(V(4,1)))*u(imin-abs(V(5,1)))*u(imin-abs(V(6,1)))*\\
+u(imin-abs(V(7,1)))*u(imin-abs(V(8,1)))*u(imin-abs(V(9,1)))*u(imin-abs(V(10,1)))-u(imax-abs(V(3,1)))*\\
+u(imax-abs(V(4,1)))*u(imax-abs(V(5,1)))*u(imax-abs(V(6,1)))*u(imax-abs(V(7,1)))*u(imax-abs(V(8,1)))*\\
+u(imax-abs(V(9,1)))*u(imax-abs(V(10,1))))\}\\
Cs s 0 1 IC=\{0.75\}\\
Raus s 0 100meg\\
.ends SCOMB8\\
\end{tabular}
\caption{\label{tbl:VCDCG} LTspice code for VCDCG and s-block.}
\end{table}

\end{document}